\documentclass[12pt]{article}%
\usepackage{amsfonts}
\usepackage{amssymb}
\usepackage{amsmath}
\usepackage{geometry}
\usepackage{graphicx}
\usepackage{setspace}
\usepackage{booktabs}
\providecommand{\U}[1]{\protect\rule{.1in}{.1in}}
\newtheorem{theorem}{Theorem}

\newtheorem{corollary}[theorem]{Corollary}

\newtheorem{lemma}[theorem]{Lemma}

\newtheorem{remark}[theorem]{Remark}

\begin{document}

\title{ }

\begin{center}
	\textit{Original Article}\bigskip
	
	{\LARGE Discrete Choice Models for Nonmonotone }
	
	{\LARGE Nonignorable Missing Data: Identification and Inference }
	
	{\Large Eric J. Tchetgen Tchetgen, Linbo Wang, BaoLuo Sun } $\ $
	
	{\large Department of Biostatistics}$,$
	
	{\large Harvard University}
	
	\textbf{Abstract}
\end{center}

\noindent Nonmonotone missing data arise routinely in empirical studies of
social and health sciences, and when ignored, can induce selection bias and
loss of efficiency. In practice, it is common to account for nonresponse under
a missing-at-random assumption which although convenient, is rarely
appropriate when nonresponse is nonmonotone. Likelihood and Bayesian missing
data methodologies often require specification of a parametric model for the
full data law, thus \textit{a priori} ruling out any prospect for
semiparametric inference. In this paper, we propose an all-purpose approach
which delivers semiparametric inferences when missing data are nonmonotone and
not at random. The approach is based on a discrete choice model (DCM) as a
means to generate a large class of nonmonotone nonresponse mechanisms that are
nonignorable. Sufficient conditions for nonparametric identification are
given, and a general framework for fully parametric and semiparametric
inference under an arbitrary DCM\ is proposed. Special consideration is given
to the case of logit discrete choice nonresponse model (LDCM) for which we
describe generalizations of inverse-probability weighting, pattern-mixture
estimation, doubly robust estimation and multiply robust estimation.


\bigskip\pagebreak

\section{\noindent Introduction}

Missing data are of common occurence in empirical research in health and
social sciences, and will often affect\ one's ability to draw reliable
inferences whether from an experimental or nonexperimental study. Non-response
can occur in sample surveys, due to dropout or non-compliance in clinical
trials, or due to data excision by error or in order to protect
confidentiality. In many practical situations, nonresponse is nonmonotone,
that is, there may be no nested pattern of missingness such that observing
variable $X_{k}$ implies that variable $X_{j}$ is also observed, for any
$j<k$. Nonmonotone missing data patterns may occur, for instance, when
individuals who dropped out of a longitudinal study re-enter at later time
points; likewise, in regression analysis nonmonotone nonresponse may occur if
the outcome or any of the regressors may be unobserved for a subset of the
sample in an arbitrary pattern. Missing data are said to be completely-at-random
(MCAR) if the nonresponse process is independent of both observed and
unobserved variables in the full data, and missing-at-random (MAR) if,
conditional on observed variables under a nonresponse pattern, the probability
of observing the pattern does not depend on unobserved variables under the
pattern (Rubin 1976; Little and Rubin 2002, Robins et al, 1994). A nonresponse
process which is neither MCAR nor MAR is said to be missing-not-at-random (MNAR).

While complete-case analysis is perhaps the most widely-used method to handle
missing data in practice, the approach is generally not recommended as it can
give biased inferences when nonresponse is not MCAR. Formal methods to
appropriately account for incomplete data include fully parametric likelihood
and Bayesian approaches (Little and Rubin 2002; Horton and Laird 1999; Ibrahim
and Chen 2000; Ibrahim et al. 2002, 2005) which are most commonly implemented
under MAR using the EM\ algorithm or via multiple imputation (MI) (Dempster et
al, 1977, Rubin 1977; Schafer 1997). Inverse probability weighting (IPW) is
another approach to account for selection bias due to missing data (Horvitz
and Thompson 1952; Robins et al. 1994; Tsiatis 2006). While IPW estimation
avoids specification of a full-data likelihood, the approach does require a
model for the nonresponse process. However, the development of general
coherent models for nonmonotone nonresponse has proved to be particularly
challenging, even under the MAR assumption; see Robins and Gill (1997) and Sun
and Tchetgen Tchetgen (2016) for two concrete proposals and further discussion.

Despite recent progress in development of MAR methodology, as argued by Gill
and Robins (1997), Robins (1997) and Little and Rubin (2002), the assumption
is generally hard to justify on substantive grounds when nonresponse is
nonmonotone. Instead, allowing for MNAR data seems particularly befitting in
the context of nonmonotone nonresponse and has received substantial attention,
particularly in the context of fully parametric models (Deltour et al. (1999),
Albert (2000), Ibrahim et al. (2001), Fairclough et al. (1998), Troxel et al
(1998), Troxel, Lipsitz \& Harrington (1998)). MNAR approaches which do not
necessarily rely on parametric assumptions have also been developed in recent
years. Notable examples include the group permutation model (GPM) of Robins
(1997) and the block conditional MAR (BCMAR) model of Zhou et al (2010).
\ Both approaches allow for non-ignorable missing data in the sense that the
nonresponse process of a given variable may depend on values of other missing
variables. However, neither BCMAR nor GPM allows the missingness probability
of a given variable to depend on the value of the variable. Based on subject
matter considerations, it is often desirable to consider non-ignorable
processes where the missingness probability of a variable depends on the
possibly unobserved value of the variable, therefore, methods for
non-ignorable missing data mechanisms beyond BCMAR\ and GPM are of interest.

In this paper, we propose a large class of non-ignorable nonmonotone
nonresponse models, which unlike BCMAR and GPM, do not \textit{a priori} rule
out the possibility that the probability of observing a given variable may
depend on the unobserved value of the variable. Our approach is based on
so-called {\normalsize discrete choice models (DCM). DCMs were first
	introduced and are predominantly used in economics and other social sciences,
	as a principled approach for generating a large class of multinomial models to
	describe discrete choice decision making under rational utility maximization.
}In this paper,\ DCMs are used for a somewhat different purpose, as a means to
generate a large class of nonmonotone nonresponse mechanisms which are
nonignorable. Sufficient conditions for nonparametric identification are
given, and a general framework for semiparametric inference under an arbitrary
DCM\ is proposed. Special consideration is given to the case of logit discrete
choice nonresponse model (LDCM). Interestingly, our identification condition
in the case of the LDCM, states that the conditional distribution of
unobserved variables given observed variables for any nonresponse pattern,
matches the corresponding conditional distribution in complete-cases. This
latter assumption is equivalent to the well-known complete-case missing value
(CCMV) restriction in the pattern mixture (PM) literature which has previously
been developed for fully likelihood-based inference (Little,1993). Therefore,
our approach provides a comprehensive treatment of semiparametric inference
for MNAR nonresponse under Little's CCMV restriction. Specifically, in
addition to reviewing Little's (i) PM likelihood approach, we describe a
generalization of (ii) inverse-probability weighting (IPW), and (iii) both
doubly robust (DR) and multiply robust (MR) estimation, which are the
nonmonotone MNAR analogues of existing results for monotone MAR nonresponse
(Tsiatis, 2006). Our doubly robust estimators combine models (i) and (ii) but
only require one of the two models to be correct. In fact, we establish that
whenever $J$ nonresponse patterns are observed, the proposed LDCM DR
estimators can be made multiply robust (more precisely $2^{J}$-robust) in the
sense that for each nonresponse pattern, valid inferences can be obtained if
one of two pattern-specific models is correctly specified but not necessarily
both. As far as we know, our paper represents the first instance of a doubly
(2$^{J}$-) robust estimator obtained for a general nonmonotone nonignorable
missing data model that is just-identified from the observed data alone. We
emphasize that our proposed inferences under the LDCM are quite attractive as
a generic nonignorable approach for arbitrary nonmonotone patterns, mainly
because they are somewhat easy to implement, have good robustness properties,
and appear to have good finite sample performance as we illustrate via
simulation studies and an HIV\ data application. In closing, we briefly
consider IPW inference for DCMs outside of the LDCM, which can generally be
used to account for nonmonotone nonignorable missing data even when Little's
CCMV\ condition fails and therefore the LDCM may not be appropriate.

\section{\noindent{\protect\Large Notation and definitions}}

\noindent Suppose full data consist of $n$ i.i.d. realizations of
{\normalsize a random }${\normalsize K}${\normalsize -vector}
{\normalsize $L=(L_{1},...,L_{K})^{\prime}$. Let $R$ denote the scalar random
	variable encoding missing data patterns, and }$J$ denote the total number of
observed patterns{\normalsize . For missing data pattern $R=r$, where $1\leq
	r\leq J\leq2^{K}$, we use $L_{(r)}\ $and }$L_{\left(  -r\right)  }$ to denote
observed and unobserved components of $L$, respectively so that
{\normalsize $L=(L_{(r)},$}$L_{\left(  -r\right)  }).$ {\normalsize We reserve
	$r=1$ to denote complete cases. Throughout, denote }$\Pr\left\{
R=r|L\right\}  =\pi_{r}(L)=\Pi_{r}$ for all $r.$ {\normalsize For each
	realization, we observe $\left(  R,L_{(R)}\right)  $. }For instance, suppose
the full data $L$ is a bivariate binary vector $(L_{1},L_{2})$ and the
following $J=3$ nonmonotone nonresponse patterns are observed in the sample:
$R=1,${\normalsize $L_{(1)}=L;R=2,$} {\normalsize $L_{(2)}=L_{1};$ and
	$R=3,L_{(3)}=L_{2}.$}

{\normalsize Throughout, we also make the following positivity assumption,
	\begin{equation}
	\Pi_{1}>\sigma>0\text{ a.s.},\label{eq:assume2}%
	\end{equation}
	for a fixed positive constant $\sigma$, that is, the probability of being a
	complete-case is bounded away from zero almost surely. Assumption
	(\ref{eq:assume2}) will be needed for nonparametric identification of the full
	data distribution, and its smooth functionals as well as finite asymptotic
	variance of IPW estimators (Robins et al, 1999). As further discussed in
	Section 2.3, complete-case IPW\ relies on obtaining a consistent estimator of
}$\pi_{1}(L)=1-\sum_{r\neq1}\pi_{r}(L)$ which in turn requires estimating the
nonresponse process $\left\{  \pi_{r}(L):r\right\}  .$ The nonresponse process
clearly fails to be nonparametrically identified under assumption
{\normalsize (\ref{eq:assume2}) only. }In the next section, we describe a set
of sufficient conditions to identify a model for the complete-case probability
$\pi_{1}(L)$ under the discrete choice framework when missingness is
nonmonotone and not at random.

Our first result provides a generic nonparametric representation of the joint
law of $f(R,L)$ that will be used throughout. The result adapts the
generalized odds ratio parametrization of a joint distribution due to Chen
(2010) to the missing data context; see also Tchetgen Tchetgen et al (2010).
\ Let $\mathrm{Odds}_{r}\left(  L\right)  =\pi_{r}\left(  L\right)  /\pi
_{1}\left(  L\right)  .$ We have the following result.

\begin{lemma}
	We have that
	\[
	f(R,L)=\frac{%
		{\displaystyle\prod\limits_{r\neq1}}
		\mathrm{Odds}_{r}\left(  L\right)  ^{I\left(  R=r\right)  }f\left(
		L|R=1\right)  }{%
		{\displaystyle\iint}
		{\displaystyle\prod\limits_{r\neq1}}
		\mathrm{Odds}_{r}\left(  l^{\ast}\right)  ^{I\left(  r^{\ast}=r\right)
		}f\left(  l^{\ast}|R=1\right)  d\mu\left(  r^{\ast},l^{\ast}\right)  },
	\]
	provided $%
	{\displaystyle\iint}
	{\displaystyle\prod\limits_{r\neq1}}
	\mathrm{Odds}_{r}\left(  l^{\ast}\right)  ^{I\left(  r^{\ast}=r\right)
	}f\left(  l^{\ast}|R=1\right)  d\mu\left(  r^{\ast},l^{\ast}\right)  <\infty,$
	with $\mu$ a dominating measure of the CDF\ of $\left(  R,L\right)  .$ \ \ \ \ \ \ \ \ \ \ \ \ \ \ \ \ \ \ \ \ \ \ \ \ \ \ \ \ \ \ \ \ \ \ \ \ \ \ \ \ \ \ \ \ \ \ \ \ \ \ \ \ \ \ \ \ \ \ \ \ \ \ \ \ \ \ \ \ \ \ \ \ \ \ \ \ \ \ \ \ \ \ \ \ \ \ \ \ \ \ \ \ \ \ \ \ \ \ \ \ \ \ \ \ \ \ \ \ \ \ \ \ \ \ \ \ \ \ \ \ \ \ \ \ \ \ \ \ \ \ \ \ \ \ \ \ \ \ \ \ \ \ \ \ \ \ \ \ \ \ \ \ \ \ \ \ \ \ \ \ \ \ \ \ \ \ \ \ \ \ \ \ \ \ \ \ \ \ \ \ \ \ \ \ \ \ \ \ \ \ \ \ \ \ \ \ \ \ \ \ \ \ \ \ \ \ \ \ \ \ \ \ \ \ \ \ \ \ \ \ \ \ \ \ \ \ \ \ \ \ \ \ \ \ \ \ \ \ \ \ \ \ \ \ \ \ \ \ \ \ 
\end{lemma}

Lemma 1 clarifies what the identification task entails, because under
assumption ${\normalsize (\ref{eq:assume2}),}$ $f\left(  L|R=1\right)  $ is
just-identified, and therefore $f(R,L)$ is nonparametrically just-identified
only if one can just-identify $\mathrm{Odds}_{r}\left(  L\right)  $ for all
$r.$ Below we describe a sufficient condition for identification under the
discrete choice model of the nonresponse process $.$

\section{\noindent Identification}

\subsection{\noindent{\protect\Large The discrete choice nonresponse model}}

\noindent{\normalsize The DCM associates with each realized nonresponse
	pattern }${\normalsize r=1,...,}$$J\leq2^{K}$ {\normalsize an underlying
	utility function }$U_{r}=\mu_{r}\left(  L\right)  +\varepsilon_{r},$ where
$\left\{  \varepsilon_{r}:r\right\}  $ are i.i.d. with cumulative distribution
function $F_{\varepsilon},$ and $\mu_{r}\left(  L\right)  $ encodes the
dependence of a person's utility on $L$ (McFadden, 1984, Train, 2009). Some
common choices of $F_{\varepsilon}$ include the extreme value distribution
(further discussed below) and the normal distribution, although in principle
any CDF could be specified. It is then assumed that a person's observed
response pattern maximizes her utility, that is $R=\arg\max_{r}\left\{
U_{r}:r\right\}  .$ Together, these assumptions imply that for each $r,$%
\begin{equation}
\Pi_{r}=\pi_{r}\left(  L\right)  =\Pr(R=r|L)=\int%
{\displaystyle\prod\limits_{s\neq r}}
F_{\varepsilon}\left(  \Delta\mu_{rs}\left(  L\right)  +\varepsilon\right)
dF_{\varepsilon}\left(  \varepsilon\right)  ,\text{ }\label{GDCM}%
\end{equation}
where $\Delta\mu_{rs}\left(  L\right)  =\mu_{r}\left(  L\right)  -\mu
_{s}\left(  L\right)  $ captures the dependence on $L$ of a difference in
utility in comparing a person's choice between nonresponse patterns $r$ and
$s,$ see Train (2009)$.$ The integral in\ $\left(  \ref{GDCM}\right)  $ is
generally not available in closed form for most choices of $F_{\varepsilon}$
(with the notable exception of the extreme value distribution, see Section
2.2), but can easily be evaluated by numerical integration using say, Gaussian
quadrature. Two interesting observations about equation $\left(
\ref{GDCM}\right)  $ are worth noting. Although not immediately apparent from
the expression in the display, equation $\left(  \ref{GDCM}\right)  $ gives
rise to a proper probability mass function, that is $\sum_{r}\pi_{r}\left(
l\right)  $ $=1$ for all values of $l$ and for any choice of $F_{\varepsilon
}.$ This remarkable result is a direct consequence of utility maximization as
a formal principle for generating multinomial probabilities $\left\{  \pi
_{r}:r\right\}  .$ A second interesting observation is that only differences
in utility matter in determining the choice probabilities; in other words, the
absolute level of a person's utility for a given nonresponse pattern is
irrelevant and only relative utility drives the choice of a nonresponse
pattern over another.\ Clearly, model $\left(  \ref{GDCM}\right)  $ is not
identifiable without an additional assumption, even given knowledge of
$F_{\varepsilon}$.

For the purpose of identification, we will consider the assumption that the
relative utility $\Delta\mu_{1r}\left(  L\right)  $ of any nonresponse pattern
$r\neq1$ compared with that of complete-case pattern $r=1$, only depends on
data observed under both patterns, that is
\begin{equation}
\Delta\mu_{1r}\left(  L\right)  =\Delta\mu_{1r}\left(  L_{(r)}\right)  \text{
	for all }r\text{ almost surely.}\label{ID}%
\end{equation}
The assumption essentially states that when faced with the choice between
nonresponse pattern $r\neq1$ versus providing complete data, the excess
utility a subject would experience choosing one over the other only depends on
data observed under both choices. Under the assumption, one may write
\begin{equation}
\Pi_{r}=\int%
{\displaystyle\prod\limits_{s\neq r}}
F_{\varepsilon}\left(  \Delta\mu_{1s}\left(  L_{\left(  s\right)  }\right)
-\Delta\mu_{1r}\left(  L_{\left(  r\right)  }\right)  +\varepsilon\right)
dF_{\varepsilon}\left(  \varepsilon\right)  \label{intgral}%
\end{equation}
Note that, even under assumption $\left(  \ref{ID}\right)  $, $\Pi_{r}$
generally depends on unobserved variables for all $r$, and therefore, data are
missing not at random, and the corresponding observed data likelihood is
nonignorable. Nevertheless, as we show in Section 5, given any continuous
$F_{\varepsilon}$, equation $\left(  \ref{intgral}\right)  $ is
nonparametrically identified for each $r$ provided $\left(  \ref{eq:assume2}%
\right)  $ holds.\ \ We leave the detailed discussion of inference under
user-specified $F_{\varepsilon}$ to Section 5, instead, to fix ideas, we
further discuss identification and inference under the logit DCM. 

\subsection{\noindent{\protect\large The logit discrete choice model}}

\noindent In the special case where $F_{\varepsilon}$ is the extreme value
distribution, the integral in equation\ $\left(  \ref{GDCM}\right)  $ is
available in closed-form, and gives the following logit DCM (Train, 2009):
$\pi_{r}\left(  L\right)  =\mathrm{Odds}_{r}\left(  L\right)  /(1+%
{\displaystyle\sum\limits_{s\neq 1}}
\mathrm{Odds}_{s}\left(  L\right)  ),$ where $\mathrm{Odds}_{r}\left(
L\right)  =\exp\left(  \Delta\mu_{1r}\left(  L\right)  \right)  $ for all $r.$
Under $\left(  \ref{ID}\right)  $, $\mathrm{Odds}_{r}\left(  L\right)
=\mathrm{Odds}_{r}\left(  L_{\left(  r\right)  }\right)  ,$ and therefore
\begin{equation}
\Pi_{r}=\frac{\mathrm{Odds}_{r}\left(  L_{\left(  r\right)  }\right)  }{1+%
	{\displaystyle\sum\limits_{s\neq 1}}
	\mathrm{Odds}_{s}\left(  L_{\left(  s\right)  }\right)  },\text{ for all
}r\neq1.\label{LDCM}%
\end{equation}
{In order to illustrate (5), briefly consider an example with }$%
L=(L_{1},L_{2},L_{3}).${\ Suppose that there are 4 nonresponse
	patterns, }$L_{(1)}=L,L_{(2)}=(L_{1},L_{2}),L_{(3)}=L_{3},L_{(4)}=%
\varnothing .${\ Then, by (3) }$\mathrm{Odds}_{2}\left( L\right) =%
\mathrm{Odds}_{2}\left( L_{(2)}\right); \mathrm{Odds}_{3}\left( L\right) =%
\mathrm{Odds}_{3}\left( L_{(3)}\right); \mathrm{Odds}_{4}\left( L\right) =%
\mathrm{Odds}_{4}\left( L_{(4)}\right) =\mathrm{Odds}_{4}${\ is a
	constant. Furthermore, according to (5) }$\Pi _{2}=\mathrm{Odds}_{2}\left(
L_{(2)}\right) /c(L);\Pi _{3}=\mathrm{Odds}_{3}\left( L_{(3)}\right) /c(L);$%
{\ }$\Pi _{4}=\mathrm{Odds}_{4}/c(L),$ where $c\left( L\right)
=\left( 1+\sum_{s\neq 1}\mathrm{Odds}_{s}\left( L_{(s)}\right) \right).$ {Therefore, by virtue of $c(L)$, the 
	nonresponse probabilities $\Pi _j$, $j=2,3,4$ are each a function of $\tilde{L}=\cup_{j=2,3,4} L_{(j)}$, the union set of observed variables across all the nonresponse patterns. Since the variable set $\tilde{L}\setminus L_{(j)}$ is not observed for each of the missing data patterns $j=2,3,4$, the nonresponse 
	process is clearly MNAR. In particular, $\Pi_4$  is a function 
	of $\tilde{L}$ even though no variable is observed in the fourth missing data pattern.}

Interestingly, an equivalent characterization of equation $\left(
\ref{LDCM}\right)  $ is:
\begin{equation}
L_{\left(  -r\right)  }|R=r,L_{\left(  r\right)  }\sim L_{\left(  -r\right)
}|R=1,L_{\left(  r\right)  }\text{ \ for all }r\neq1,\label{CCMV}%
\end{equation}
which states that the conditional distribution of unobserved variables
$L_{(-r)}$ given observed variables $L_{(r)}$ for nonresponse pattern $r$
matches the corresponding conditional distribution among complete-cases.
Although the LDCM is derived as a particular DCM, one could in principle take
$\left(  \ref{CCMV}\right)  $ as primitive identifying condition without
necessarily making reference to a DCM\ and the existence of its associated
variables $\left\{  \varepsilon_{r}:r\right\}  .$This amounts to nonparametric
identification under the complete-case missing value restriction of Little
(1993). As shown in Section 5, adoption of the more general DCM\ framework is
advantageous as it gives rise to a richer class of nonresponse models and
facilitates identification; in fact, a different choice for the distribution
$F_{\varepsilon}$ corresponds to a nonmonotone not at random nonresponse model
which does not generally satisfy Little's CCMV restriction but is nevertheless
just-identified under $\left(  \ref{eq:assume2}\right)  $ and $\left(
\ref{ID}\right)  $. \ 

It is instructive to compare condition $\left(  \ref{CCMV}\right)  $ to
standard MAR, which states that
\begin{equation}
L_{\left(  -r\right)  }|R=r,L_{\left(  r\right)  }\sim L_{\left(  -r\right)
}|L_{\left(  r\right)  }\text{ \ for all
}r\neq1,\label{MAR}%
\end{equation}
i.e. the conditional distribution for pattern $r$ matches the conditional
distribution obtained upon marginalizing across all nonresponse patterns.
Clearly, conditions $\left(  \ref{CCMV}\right)  $ and $\left(  \ref{MAR}%
\right)  $ have fundamentally different implications for inference.
Specifically, it is well known that when the nonresponse process and the full
data distribution depend on separate parameters, the MAR\ assumption implies
that the part of the observed data likelihood which depends on the full data
parameter factorizes from the nonresponse process. The missing data mechanism
is then said to be ``ignorable'' (Little and Rubin, 2002) because it is possible
to learn about the full data law without necessarily estimating the missing
data process, or equivalently, it is possible to learn about the missing data
process without modeling the full data law (Sun and Tchetgen Tchetgen, 2016).
No such factorization is in general available under CCMV as the missing data
process is nonignorable. In spite of possible challenges due to lack of
factorization, as shown later in the paper, estimation of nonmonotone
non-response mechanisms under $\left(  \ref{CCMV}\right)  $ is nevertheless
relatively straightforward. \ Furthermore, assumption $\left(  \ref{CCMV}%
\right)  $ is invariant to the number and nature of other nonresponse patterns
potentially realized in the observed data. In contrast, MAR does not enjoy a
similar invariance property because addition or deletion of a nonresponse
pattern from the observed sample changes the interpretation of $\left(
\ref{MAR}\right)  $ as it implies marginalizing over a different set of
nonresponse patterns to obtain the right-hand side of equation $\left(
\ref{MAR}\right)  $. \ Finally, note that assumptions $\left(  \ref{CCMV}%
\right)  $ and $\left(  \ref{MAR}\right)  $ only coincide when there is a
single nonresponse pattern, i.e. $J=2$.

\begin{remark}
	Sun and Tchetgen Tchetgen (2016) recently proposed an approach tailored specifically to model a nonmonotone nonresponse process under MAR restriction (7). However, they did not consider the MNAR restriction (3). As restrictions (3) and (7) 	differ, the approach proposed by Sun and Tchetgen Tchetgen (2016) cannot be used 	under restriction (3).
\end{remark}

\begin{lemma}
	Suppose that assumptions $\left(  {\normalsize \ref{eq:assume2}}\right)  $ and
	$\left(  \ref{GDCM}\right)  $ hold with $F_{\varepsilon}$ being the extreme
	value distribution, then if $\left(  \ref{ID}\right)  $ holds, the joint
	distribution $f(R,L)$ is nonparametrically just-identified from the observed
	data $\left(  L_{R},R\right)  ,$ with%
	\begin{equation}
	f(R,L)=\frac{%
		{\displaystyle\prod\limits_{r\neq1}}
		\mathrm{Odds}_{r}\left(  L_{\left(  r\right)  }\right)  ^{I\left(  R=r\right)
		}f\left(  L|R=1\right)  }{%
		{\displaystyle\iint}
		{\displaystyle\prod\limits_{r\neq1}}
		\mathrm{Odds}_{r}\left(  l_{\left(  r\right)  }^{\ast}\right)  ^{I\left(
			r^{\ast}=r\right)  }f\left(  l^{\ast}|R=1\right)  d\mu\left(  r^{\ast}%
		,l^{\ast}\right)  }, \label{jointLIk}%
	\end{equation}
	where $\mu$ is a dominating measure of the CDF\ of $\left(  R,L\right)  .$
\end{lemma}

Lemma 2 gives an explicit expression for\ $f(R,L)$ which appears to be new,
and can be used to compute the full data density $f(L)=\sum_{r}f(r,L).$ In
addition, equation $(8)$ can be used for maximum likelihood estimation.
Specifically, let $f\left(  L|R=1;\mathbf{\eta}\right)  $ denote a parametric
model for $f\left(  L|R=1\right)  $ with unknown parameter $\mathbf{\eta}.$
Likewise, consider a parametric model for nonresponse process $\Pi_{r}\left(
\mathbf{\alpha}\right)  =\mathrm{Odds}_{r}\left(  L_{\left(  r\right)
};\alpha_{r}\right)  /\{1+%
{\displaystyle\sum\limits_{s\neq 1}}
\mathrm{Odds}_{s}\left(  L_{\left(  s\right)  };\alpha_{s}\right)  \}$ with
unknown parameter $\mathbf{\alpha=\{}\alpha_{r}:r\}$, where $\alpha_{r}$
indexes a parametric model for $\mathrm{Odds}_{r}\left(  L_{\left(  r\right)
};\alpha_{r}\right)  .$ Let $f(R,L;\mathbf{\theta})$ denote the corresponding
model for $f(R,L),$ where $\mathbf{\theta}=\left(  \mathbf{\eta,\alpha
}\right)  .$ The maximum likelihood estimator (MLE) $\mathbf{\hat{\theta}%
}_{mle}$ maximizes the observed data log-likelihood $\mathbb{P}_{n}\log\int
f(R,L;\mathbf{\theta})d\mu\left(  L_{(-R)}\right),$ where $\mathbb{P}%
_{n}\left(  \cdot\right)  =n^{-1}\sum_{i}\left(  \cdot\right)  _{i}.$ The full
data likelihood $f(L;\mathbf{\hat{\theta}}_{mle})=\int f(r,L;\mathbf{\hat
	{\theta}}_{mle})d\mu\left(  r\right)  $ can then be used to make inferences
about a given full data functional of interest according to the plug-in
principle. By standard likelihood theory, the MLE is asymptotically efficient
in the model $\mathcal{M}_{lik}$ corresponding to the set of laws $\left\{
f(R,L;\mathbf{\theta}):\mathbf{\theta}\right\}  $. A major drawback of maximum
likelihood inference is lack of robustness to model mis-specification, because
$\mathbf{\hat{\theta}}_{mle}$ is likely inconsistent if either $\Pi_{r}\left(
\mathbf{\alpha}\right)  $ or $f\left(  L|R=1;\mathbf{\eta}\right)  $ is
incorrectly specified. Below, we consider four semiparametric estimators which
are potentially more robust than direct likelihood maximization.

\section{Semiparametric Inference}

\subsection{Inverse-probability weighting estimation}

Suppose the parameter of interest $\beta_{0}$ is the unique solution to the
full data population estimating equation $E\left\{  U(L;\beta_{0})\right\}
=0,$ where expectation is taken over the distribution of the complete data
$L$. Note that in principle, no further restriction on the distribution of $L$
is strictly required; in fact, estimation is possible under certain weak
regularity conditions (van der Vaart, 1998) as long as a full data unbiased
estimating function exist. In the presence of missing data, the estimating
function can only be evaluated for complete-cases, who might be highly
selected even under MAR. This motivates the use of IPW estimating functions of
complete-cases to form the following complete-case population estimating
equation%
\begin{equation}
E\left\{  \frac{1\left(  R=1\right)  }{\Pi_{1}}U(L;\beta_{0})\right\}  =0,
\label{ipwpop}%
\end{equation}
which holds by straightforward
iterated expectations. We note that the IPW estimator $\widehat{\beta}_{ipw}$
which solves the empirical version of this equation will in general be
inefficient especially when the fraction of complete-cases is relatively
small, since incomplete cases are discarded (except when estimating $\Pi
_{1}).$ In the next section we will describe a strategy to recover information
from incomplete-cases by augmenting estimating function shown in equation
$\left(  \ref{ipwpop}\right)  $ to gain efficiency and potentially robustness.
The IPW estimating equations framework encompasses a great variety of settings
under which investigators may wish to account for non-monotone missing data.
These include IPW of the full data score equation, where the score function is
such an unbiased estimating function, given a model $f(L;\beta_{0})$ for the
law of the full data, in which case $\left(  \ref{ipwpop}\right)  $ reduces to
$E\left\{  1\left(  R=1\right)  \partial\log\left.  f(L;\beta)/\partial
\beta\right\vert _{\beta_{0}}/\Pi_{1}\right\}  =0$

We now describe a straightforward approach to obtain a consistent estimator of
$\Pi_{1}$ in the semiparametric model which specifies a parametric LCDM
$\left\{  \Pi_{r}\left(  \mathbf{\alpha}\right)  :r\right\}  ,$ but allows
$f\left(  L|R=1\right)  $ to remain unrestricted. We denote this model
$\mathcal{M}_{R}.$ The approach follows from the fact that $\left(
\ref{LDCM}\right)  $ implies that$:$
\[
\Pr\left(  R=r|L,R\in\left\{  1,r\right\}  \right)  =\Pi_{r,c}=\frac
{\mathrm{Odds}_{r}\left(  L_{\left(  r\right)  }\right)  }{1+\mathrm{Odds}%
	_{r}\left(  L_{\left(  r\right)  }\right)  },\text{ for all }r;
\]
which also gives the following equivalent representation of the CCMV
restriction:%
\[
R\perp\!\!\!\perp L_{\left(  -r\right)  }|R\in\left\{  r,1\right\}
\text{,}L_{\left(  r\right)  }\text{ for each }r.
\]
Note that $L_{\left(  r\right)  }$ is fully observed for observations
$R\in\left\{  1,r\right\}  $. Thus, in order to estimate the parametric model
$\left\{  \Pi_{r,c}\left(  \mathbf{\alpha}\right)  :r\right\}  ,$ for each
nonresponse pattern $r$ one may fit the following logistic regression
$\Pi_{r,c}\left(  \alpha_{r}\right)  =\mathrm{Odds}_{r}\left(  L_{\left(
	r\right)  };\alpha_{r}\right)  /\{1+\mathrm{Odds}_{r}\left(  L_{\left(
	r\right)  };\alpha_{r}\right)  \}$ by maximum likelihood estimation restricted
to the subset of data containing complete-cases and incomplete-cases of
pattern $r$ only. Thus, we define the restricted MLE%
\begin{align*}
\widetilde{\alpha}_{r} &  =\underset{\alpha_{r}}{\arg\max}\mathbb{P}%
_{n}\text{\textrm{llik}}_{r,c}\left(  \alpha_{r}\right)  \\
&  =\underset{\alpha_{r}}{\arg\max}\mathbb{P}_{n}\left\{  I\left(  R=r\right)
\log\Pi_{r,c}\left(  \alpha_{r}\right)  +I\left(  R=1\right)  \log\left(
1-\Pi_{r,c}\left(  \alpha_{r}\right)  \right)  \right\}  .
\end{align*}
Under assumption {\normalsize (\ref{eq:assume2}),} the restricted MLE
$\widetilde{\mathbf{\alpha}}$ is consistent and asymptotically normal under
model $\mathcal{M}_{R}.$ The resulting estimator of the complete-case
probability $\Pi_{1}$ under\ $\mathcal{M}_{R}$ is $$\Pi_{1}\left(
\widetilde{\mathbf{\alpha}}\right)= \frac{1}{1+\sum_{s \neq 1}\mathrm{Odds}_{s}\left(L_{(s)};\widetilde{\mathbf{\alpha}}_s\right)},$$ which in turn, provides the IPW estimator
$\widehat{\beta}_{ipw}$ of $\beta$ which solves%
\begin{equation}
\mathbb{P}_{n}\left\{  U_{ipw}(L_{(R)},R;\widehat{\beta}_{ipw}%
,\widetilde{\mathbf{\alpha}})\right\}  =0,\label{ipwee}%
\end{equation}
\noindent where $U_{ipw}(L_{(R)},R;\widehat{\beta}_{ipw}%
,\widetilde{\mathbf{\alpha}})=1\left(  R=1\right)  U(L;\widehat{\beta}%
_{ipw})/\Pi_{1}\left(  \widetilde{\mathbf{\alpha}}\right)  $. Under standard
regularity conditions, one can show that under $\mathcal{M}_{R}$ the IPW
estimator $\widehat{\beta}_{ipw}$ will in large sample be approximately normal
with mean $\beta_{0}$ and asymptotic variance $\hat{\Gamma}_{ipw}^{-1}%
\hat{\Omega}_{ipw}\hat{\Gamma}_{ipw}^{-1},$ where%
\begin{align*}
\hat{\Gamma}_{ipw}^{-1} &  =-\left.  \frac{\partial}{\partial\beta^{T}%
}\mathbb{P}_{n}\left\{  U_{ipw}(L_{(R)},R;\beta,\widetilde{\mathbf{\alpha}%
})\right\}  \right\vert _{\widehat{\beta}_{ipw}};\\
\hat{\Omega}_{ipw} &  =n^{-1}\mathbb{P}_{n}\left\{  \left[  U_{ipw}%
(L_{(R)},R;\widehat{\beta}_{ipw},\widetilde{\mathbf{\alpha}})+\left.
\frac{\partial}{\partial\mathbf{\alpha}^{T}}\mathbb{P}_{n}\left\{
U_{ipw}(L_{(R)},R;\widehat{\beta}_{ipw},\mathbf{\alpha})\right\}  \right\vert
_{\widetilde{\mathbf{\alpha}}}\widehat{IF}_{\mathbf{\alpha}}\right]
^{\otimes2}\right\}  ;\\
\widehat{IF}_{\mathbf{\alpha}} &  =-\left[  \left.  \frac{\partial^{2}%
}{\partial\mathbf{\alpha}\partial\mathbf{\alpha}^{T}}\mathbb{P}_{n}\left\{
\sum_{r\neq1}\text{\textrm{llik}}_{r,c}\left(  \alpha_{r}\right)  \right\}
\right\vert _{\widetilde{\mathbf{\alpha}}}\right]  ^{-1}\left.  \frac
{\partial}{\partial\mathbf{\alpha}}\left\{  \sum_{r\neq1}\text{\textrm{llik}%
}_{r,c}\left(  \alpha_{r}\right)  \right\}  \right\vert
_{\widetilde{\mathbf{\alpha}}}.
\end{align*}
For inference about a component of $\beta_0$, one may report the corresponding Wald-type 95\% confidence interval.

\subsection{Pattern-mixture LDCM estimation}

In this Section, we consider an alternative approach for obtaining inferences
about the full data parameter $\beta_{0}$ defined in the previous Section. The
approach is a slight generalization of the well-known pattern-mixture approach
due to Little (1993). To proceed, note that%
\begin{align}
E\left\{  U(L;\beta_{0})\right\}   &  =E\left[  E\left\{  U(L;\beta
_{0})|R,L_{\left(  R\right)  }\right\}  \right]  ,\nonumber\\
&  =E\left[  E\left\{  U(L;\beta_{0})|R=1,L_{\left(  R\right)  }\right\}
\right]  \nonumber\\
&  =E\left[  \sum_{r}I(R=r)E\left\{  U(L;\beta_{0})|R=1,L_{\left(  r\right)
}\right\}  \right]  \label{PMpop}\\
&  =0\nonumber
\end{align}
\noindent where the second equality follows from $\left(  \ref{CCMV}\right)
.$ Now, consider the semiparametric model $\mathcal{M}_{L}$ which posits
parametric model $f\left(  L|R=1;\mathbf{\eta}\right)  $ while allowing the
nonresponse process $\left\{  \Pi_{r}:r\right\}  $ to remain unrestricted. Let
$\widetilde{\mathbf{\eta}}$ denote the restricted MLE of $\mathbf{\eta}$
$\mathbf{\,}$in $\mathcal{M}_{L}$ obtained using only complete-case data, i.e.
$\widetilde{\eta}=\underset{\mathbf{\eta}}{\arg\max}\mathbb{P}_{n}%
$\textrm{llik}$_{l,c}\left(  \eta\right)  =\underset{\mathbf{\eta}}{\arg\max
}\mathbb{P}_{n}I\left(  R=1\right)  \log f\left(  L|R=1;\mathbf{\eta}\right)
.$ An empirical version of equation $\left(  \ref{PMpop}\right)  $ can then be
used to obtain the following pattern mixture estimator $\widehat{\beta}_{pm}$
of $\beta_{0},$%
\begin{equation}
0=\mathbb{P}_{n}\left[  U_{pm}(L_{(R)},R;\widehat{\beta}_{pm},\widetilde{\eta
})\right]  ,\label{pmee0}%
\end{equation}
where
\begin{equation}
U_{pm}(L_{(R)},R;\widehat{\beta}_{pm},\widetilde{\eta})=\sum_{r}%
I(R=r)E\left\{  U(L;\widehat{\beta}_{pm})|R=1,L_{\left(  r\right)
};\widetilde{\eta}\right\}  ,\label{pmee}%
\end{equation}
and $E\left\{  U(L;\widehat{\beta}_{pm})|R=1,L_{\left(  r\right)
};\widetilde{\eta}\right\}  =\int U(l_{\left(  -r\right)  },L_{(r)}%
;\widehat{\beta}_{pm})f\left(  l_{\left(  -r\right)  }|L_{(r)}%
|R=1;\widetilde{\eta}\right)  d\mu\left(  l_{\left(  -r\right)  }\right)  $.
\ Note that in order to ensure that models$\left\{  f\left(  l_{\left(
	-r\right)  }|L_{\left(  r\right)  }|R=1;\widetilde{\eta}\right)
,r\neq1\right\}  $ are compatible, one may need to specify a model for
$f\left(  L|R=1\right)  ;$ this is effectively the approach followed by Little
(1993). Also note that in the pattern mixture approach, the model for $f(L)$
which is of primary scientific interest is indirectly specified via models for
the various conditional densities $\left\{  f\left(  l_{\left(  -r\right)
}|L_{(r)}|R=1\right)  ,r\neq1\right\}  $ and the marginal densities $\left\{
f\left(  L_{\left(  r\right)  }|R=r\right)  ,r\neq1\right\}  $ according to
the following mixture: $f(L)=\sum_{r}f\left(  l_{\left(  -r\right)
}|L_{\left(  r\right)  }|R=1\right)  f\left(  l_{\left(  r\right)
}|R=r\right)  \Pr(R=r)$ (Little, 1993). Under standard regularity conditions,
one can show that in large samples, $\widehat{\beta}_{pm}$ will be
approximately normal with mean $\beta_{0\text{ }}$and asymptotic variance
consistently estimated by $\hat{\Gamma}_{pm}^{-1}\hat{\Omega}_{pm}\hat{\Gamma
}_{pm}^{-1}$ where%
\begin{align*}
\hat{\Gamma}_{pm}^{-1} &  =-\left.  \frac{\partial}{\partial\beta^{T}%
}\mathbb{P}_{n}\left\{  U_{pm}(L_{(R)},R;\beta,\widetilde{\eta})\right\}
\right\vert _{\widehat{\beta}_{pm}};\\
\hat{\Omega}_{pm} &  =n^{-1}\mathbb{P}_{n}\left[  U_{pm}(L_{(R)}%
,R;\widehat{\beta}_{pm},\widetilde{\eta})+\left.  \frac{\partial}%
{\partial\mathbf{\eta}^{T}}\mathbb{P}_{n}U_{pm}(L_{(R)},R;\widehat{\beta}%
_{pm},\mathbf{\eta})\right\vert _{\widetilde{\mathbf{\eta}}}\widehat{IF}%
_{\mathbf{\eta}}\right]  ^{\otimes2};\\
\widehat{IF}_{\mathbf{\eta}} &  =-\left[  \left.  \frac{\partial^{2}}%
{\partial\eta\partial\eta^{T}}\mathbb{P}_{n}\left\{  \text{\textrm{llik}%
}_{l,c}\left(  \eta\right)  \right\}  \right\vert _{\widetilde{\mathbf{\eta}}%
}\right]  ^{-1}\left.  \frac{\partial}{\partial\eta}\left\{  \sum_{r\neq
1}\text{\textrm{llik}}_{l,c}\left(  \eta\right)  \right\}  \right\vert
_{\widetilde{\mathbf{\eta}}}.
\end{align*}

\subsection{Doubly robust and multiply robust LDCM estimation}

We have now described two separate approaches for estimating the full data
functional $\beta_{0\text{ }}$under the LDCM, IPW\ and PM estimation, each of
which depends on a separate part (i.e. variation independent parameter) of the
joint distribution of $f\left(  R,L\right)  $ given in Lemma 2. As previously
discussed, validity of IPW estimation relies on correct specification of the
nonresponse model $\mathcal{M}_{R},$ while PM estimation relies for
consistency on correct specification of $\mathcal{M}_{L}.$ Because when $L$ is
sufficiently high dimensional, one cannot be confident that either, if any,
model is correctly specified, it is of interest to develop a doubly robust
estimation approach, which is guaranteed to deliver valid inferences about
$\beta_{0\text{ }}$provided that either $\mathcal{M}_{R}$ or $\mathcal{M}_{L}$
is correctly specified, but not necessarily both. That is, we aim to develop a
consistent estimator of $\beta_{0\text{ }}$in the semiparametric union model
$\mathcal{M}_{DR}=$ $\mathcal{M}_{R}$ $\cup\mathcal{M}_{L}.$

In order to describe the DR approach, let
\begin{align*}
V\left(  \beta,\mathbf{\alpha,\eta}\right)   &  \equiv v\left(  L_{(R)}%
,R;\beta,\mathbf{\alpha,\eta}\right) \\
&  =\left\{  \frac{1\left(  R=1\right)  }{\Pi_{1}\left(  \mathbf{\alpha
	}\right)  }U(L;\beta)\right\} \\
&  -\frac{1\left(  R=1\right)  }{\Pi_{1}\left(  \mathbf{\alpha}\right)  }%
\sum_{r\neq1}\Pi_{r}\left(  \mathbf{\alpha}\right)  E\left[  U(L;\beta
)|L_{\left(  r\right)  },R=1;\mathbf{\eta}\right] \\
&  +\sum_{r\neq1}I\left(  R=r\right)  E\left[  U(L;\beta)|L_{\left(  r\right)
},R=1;\mathbf{\eta}\right]
\end{align*}
and let $\widehat{\beta}_{dr}$ denote the solution to the equation
\begin{equation}
0=\mathbb{P}_{n}V\left(  \widehat{\beta}_{dr},\widetilde{\mathbf{\alpha}%
}\mathbf{,}\widetilde{\mathbf{\eta}}\right)  . \label{dree}%
\end{equation}
We have the following result.

\begin{theorem}
	Suppose that assumptions $\left(  {\normalsize \ref{eq:assume2}}\right)  $ and
	$\left(  \ref{GDCM}\right)  $ hold with $F_{\varepsilon}$ the extreme value
	distribution. Then, under standard regularity conditions, we have that
	$\widehat{\beta}_{dr}$ is consistent and asymptotically normal in the union
	model $\mathcal{M}_{DR}$ with asymptotic variance consistently estimated by
	$\hat{\Gamma}_{dr}^{-1}\hat{\Omega}_{dr}\hat{\Gamma}_{dr}^{-1},$ where%
	\begin{align*}
	\hat{\Gamma}_{dr}^{-1}  &  =-\left.  \frac{\partial}{\partial\beta^{T}%
	}\mathbb{P}_{n}\left\{  V\left(  \beta,\widetilde{\mathbf{\alpha}}%
	\mathbf{,}\widetilde{\mathbf{\eta}}\right)  \right\}  \right\vert
	_{\widehat{\beta}_{dr}};\\
	\hat{\Omega}_{dr}  &  =n^{-1}\mathbb{P}_{n}\left[  V\left(  \widehat{\beta
	}_{dr},\widetilde{\mathbf{\alpha}}\mathbf{,}\widetilde{\mathbf{\eta}}\right)
	\right. \\
	&  \left.  +\left.  \frac{\partial}{\partial\mathbf{\eta}^{T}}\mathbb{P}%
	_{n}\left\{  V\left(  \widehat{\beta}_{dr},\widetilde{\mathbf{\alpha}%
	}\mathbf{,\eta}\right)  \right\}  \right\vert _{\widetilde{\mathbf{\eta}}%
}\widehat{IF}_{\mathbf{\eta}}+\left.  \frac{\partial}{\partial\mathbf{\alpha
}^{T}}\mathbb{P}_{n}\left\{  V\left(  \widehat{\beta}_{dr},\mathbf{\alpha
,}\widetilde{\mathbf{\eta}}\right)  \right\}  \right\vert
_{\widetilde{\mathbf{\alpha}}}\widehat{IF}_{\mathbf{\alpha}}\right]
^{\otimes2}.
\end{align*}
\end{theorem}

The above theorem formally establishes the DR property of $\widehat{\beta
}_{dr}$. Instead of the above estimators of asymptotic variance, one may use
the nonparametric bootstrap to obtain inferences based on either
$\widehat{\beta}_{dr},$ $\widehat{\beta}_{ipw}$ or $\widehat{\beta}_{pm}.$

\begin{remark}
	      Equation $\left(  \ref{jointLIk}\right)  $ of Lemma 2 implies that $f\left(
	R=1|l\right)  $ (which only depends on 
	$\{\mathrm{Odds}_{r}\left(  l_{\left(
		r\right)  }\right) :r\} $) and $f(l|R=1)$ are variation independent under the CCMV
	restriction. This variation independence is important as double robustness is
	meaningful only if it is possible \emph{a priori }for both of the nuisance
	models to be correctly specified, see Robins and Rotnitzky (2001) and
	Richardson et al (2016, Remark 3.1). Note however, that in general $f(l|r)$
	and $f(r|l)$ are variation dependent even under CCMV. 
\end{remark}

Interestingly, it is possible to make the estimator $\widehat{\beta}_{dr}$
even more robust by the following modification to estimation of the nuisance
parameter $\mathbf{\eta}$. Specifically, suppose that for each $r,$ the
conditional density $f\left(  L_{(-r)}|L_{\left(  r\right)  },r;\mathbf{\eta
}\right)  =f\left(  L_{(-r)}|L_{\left(  r\right)  },r;\eta_{r}\right)
=f\left(  L_{(-r)}|L_{\left(  r\right)  },R=1;\eta_{r}\right)  $ only depends
on the subset of parameter $\eta_{r}\subset\mathbf{\eta,}$ where there may be
parameter overlap across patterns $\eta_{r}\cap\eta_{r^{\prime}}%
\neq\varnothing$ for distinct patterns $r$ and $r^{\prime}.$ Let
$\mathcal{M}_{L}\left(  r\right)  $ denote the semiparametric model which only
specifies $f\left(  L_{(-r)}|L_{\left(  r\right)  },R=1;\eta_{r}\right)
,~$allowing the density of $f(L_{(r)}|R=1)$ and the missing data process to
remain unspecified. Note that $\mathcal{M}_{L}\subseteq%
{\displaystyle\bigcap\limits_{r\neq1}}
\mathcal{M}_{L}\left(  r\right)  $. Let $\overline{\eta}_{r}$ denote the
complete-case MLE under $\mathcal{M}_{L}\left(  r\right)  :$ $\overline{\eta
}_{r}=\underset{\eta_{r}}{\arg\max}\mathbb{P}_{n}I\left(  R=1\right)
f(L_{(-r)}|L_{\left(  r\right)  },R=1;\eta_{r}).$ Likewise, let $\mathcal{M}%
_{R}\left(  r\right)  $ denote the semiparametric model that specifies the
nonresponse model $\Pi_{r,c}\left(  \alpha_{r}\right)  ,$ and is otherwise
unspecified. Note that $\mathcal{M}_{R}=%
{\displaystyle\bigcap\limits_{r\neq1}}
\mathcal{M}_{R}\left(  r\right)  .$ Consider the following pattern-specific
union model $\mathcal{M}_{DR}\left(  r\right)  =\mathcal{M}_{R}\left(
r\right)  \cup\mathcal{M}_{L}\left(  r\right)  ,$ which is the set of laws
with either $\mathcal{M}_{R}\left(  r\right)  $ or $\mathcal{M}_{L}\left(
r\right)  $ correctly specified. The intersection submodel of these laws
$\mathcal{M}_{MR}=%
{\displaystyle\bigcap\limits_{r\neq1}}
\mathcal{M}_{DR}\left(  r\right)  =%
{\displaystyle\bigcap\limits_{r\neq1}}
$ $\left\{  \mathcal{M}_{R}\left(  r\right)  \cup\mathcal{M}_{L}\left(
r\right)  \right\}  $ is the set of laws such that the union model for each
$r$ holds. Note that $\mathcal{M}_{DR}\subseteq\mathcal{M}_{MR}$ since the
first union model requires that either the entire nonresponse process is
correctly specified, i.e.$\
{\displaystyle\bigcap\limits_{r\neq1}}
\mathcal{M}_{R}\left(  r\right)  $ holds$,$ or the joint complete-case
distribution of $L$ is correctly specified, i.e. $%
{\displaystyle\bigcap\limits_{r\neq1}}
\mathcal{M}_{L}\left(  r\right)  $ holds; in contrast, $\mathcal{M}_{MR}$
requires only correct specification of one of the two models for each pattern.
An estimator of $\beta_{0}$ that is consistent in model $\mathcal{M}_{MR}$ is
said to be multiply-robust, or more precisely $2^{J}-$robust (Vansteelandt et
al, 2007) for a $J$ non-monotone missing data patterns. We have the following result:

\begin{corollary}
	Suppose that assumptions $\left(  {\normalsize \ref{eq:assume2}}\right)  $ and
	$\left(  \ref{GDCM}\right)  $ hold with $F_{\varepsilon}$ the extreme value
	distribution. Then, under standard regularity conditions, we have that
	$\widehat{\beta}_{mr}$ is consistent and asymptotically normal in the union
	model $\mathcal{M}_{MR}$, where $\widehat{\beta}_{mr}$ is defined as
	$\widehat{\beta}_{dr}$ with $\overline{\eta}_{r}$ used to estimate $\eta_{r}.$
\end{corollary}

The above corollary describes an estimator with the MR property which states
that given $J$ nonresponse patterns, the analyst would in principle have
(under our identifying assumptions) $2^{J}$ opportunities to obtain valid
inferences about $\beta_{0}.$ This is to be contrasted with the single chance
to valid inferences offered by IPW or PM approaches respectively, or the two
chances offered by the DR estimator. For inference, one may readily adapt the
large sample variance estimator given in Theorem 3, or alternatively use the
nonparametric bootstrap.

\subsection{Simulation Study}

We perfomed a simulation study to investigate the performance of the various
estimators described above in finite sample. \ We generated $1000$\ samples of
size $n=2000.$ We implemented the following data generating mechanism.
Independent and identically distributed $(Y,X)$ is generated from a normal mixture models: 
$(Y,X) \sim \sum\limits_{k=1}^3 \pi_k N(\mu_k, \Sigma),$ where $\pi_1 =1/2, \pi_2 = e/(2+2e), \pi_3 = 1/(2+2e), \mu_1 = (0,0)^T, \mu_2 = (1,1)^T, \mu_3 = (1,2)^T$ and 
$
\Sigma = (\sigma_{ij}), 
$ where $\sigma_{11} = \sigma{12}=1, \sigma_{22}=2.$ 
We consider four missing data patterns $L_{(R)}$: $L_{(1)}=L,$ $L_{\left(  2\right)  }=X,$ $L_{\left(
	3\right)  }=Y,$ $L_{\left(  4\right)  }=\varnothing.$ 
Conditional on the generated full data, the missing data pattern is then  generated under the following mechanism: 
\begin{flalign*}
P(R=1\mid X,Y) &= \dfrac{1}{1+\exp(X) + \exp(2Y) + \exp(-1)};\\
P(R=2\mid X,Y) &= \dfrac{\exp(X) }{1+\exp(X) + \exp(2Y) + \exp(-1)};\\
P(R=3\mid X,Y) &= \dfrac{\exp(2Y) }{1+\exp(X) + \exp(2Y) + \exp(-1)};\\
P(R=4\mid X,Y) &= \dfrac{\exp(-1) }{1+\exp(X) + \exp(2Y) + \exp(-1)}.
\end{flalign*}
Since for each missing data pattern $r$, $P(R=r\mid X,Y)$ depend on all the full data $(X,Y),$ the missing data mechanism is MNAR.
The identifiability of  normal mixture models in the MNAR setting has previously been considered in  Miao et al. (2016).  The full data target parameter of interest is
$\beta=E\left(  Y\right)  =\sum_{r}p_{r}E\left[  Y|R=r\right]  =(2+\exp
(1))/(2+2\exp(1)),$ with full data estimating equation $U\left(  \beta\right)
=Y-\beta.$

We implemented Little's PM approach as well as our IPW and DR estimators. In
doing so, correct specification of the nonresponse process entailed matching
the data generating mechanism described above, i.e. $\mathrm{Odds}_{2}\left(
L_{\left(  2\right)  }\right)  =\alpha_{20}+\alpha_{21}X,$ $\mathrm{Odds}%
_{3}\left(  L_{\left(  3\right)  }\right)  =\alpha_{30}+\alpha_{31}Y,$
$\mathrm{Odds}_{4}\left(  L_{\left(  4\right)  }\right)  =\alpha_{40}.$
Misspecification of these models occured by instead fitting $\mathrm{Odds}%
_{2}\left(  L_{\left(  2\right)  }\right)  =\alpha_{20}+\alpha_{21}X^{2}$ and
$\mathrm{Odds}_{3}\left(  L_{\left(  3\right)  }\right)  =\alpha_{30}%
+\alpha_{31}Y^{2}.$ Likewise, correct specification for the PM approach
entailed defining $E\left(  Y|R=2,X\right)  =E\left(  Y|R=1,X\right)
=\gamma_{20}+\gamma_{21}X,$ while the incorrect model $E\left(
Y|R=1,X\right)  =\gamma_{20}+\gamma_{21}X^{2}$ was used to assess the impact
of model mis-specification of the complete-case distribution. \ Note that as
$U\left(  \beta\right)  $ does not depend on $X,E\left[  U\left(
\beta\right)  |R=3,L_{(3)}\right]  =U\left(  \beta\right)  .$ We explored four
scenarios corresponding to (1)\ correct $f\left(  R|L\right)  $ and
$f(L|R=1),(2)$ correct $f\left(  R|L\right)  $ but incorrect $f(L|R=1)$; $(3)$
correct $f(L|R=1)$ but incorrect $f\left(  R|L\right)  $; finally $(4)$
incorrect $f\left(  R|L\right)  $ and $f(L|R=1)$.%
\begin{table}
	\begin{center}
		\caption{Monte Carlo results of the IPW, PM and DR estimators: accuracy of standard deviation estimator and coverage probabilities. The sample size is 2000}
		\label{tab:est}
				\vspace{10pt}
		\begin{tabular}{rcccc}
			\toprule
			& \multicolumn{1}{c}{\tt bth$^*$} & \multicolumn{1}{c}{\tt nrm} &   \multicolumn{1}{c}{\tt ccm} & \multicolumn{1}{c}{\tt bad} \\
			\midrule
			\multicolumn{4}{l}{Estimated SD / Monte Carlo SD}	 &  \\[3pt]
			\quad \quad 	\quad \quad	IPW & 0.951 & 0.951 & 0.438 & 0.438 \\ 
			PM & 0.993 & 0.979 & 0.993 & 0.979 \\ 
			DR & 0.995 & 0.995 & 0.886 & 0.725 \\[12pt]
			\multicolumn{4}{l}{Estimated SD / Bootstrapped SD }	 & \\[3pt]
			IPW & 0.994 & 0.994 & 0.932 & 0.932 \\ 
			PM & 1.000 & 1.002 & 1.000 & 1.002 \\ 
			DR & 0.999 & 0.990 & 0.973 & 0.951 \\ [12pt]
			\multicolumn{4}{l}{Coverage**}	 & \\[3pt]
			IPW & 0.938 & 0.938 & 0.080 & 0.080 \\ 
			PM & 0.954 & 0.001 & 0.954 & 0.001 \\ 
			DR & 0.948 & 0.947 & 0.953 & 0.030 \\ 
			\bottomrule
		\end{tabular}
	\end{center}	
	\footnotesize{*: {\tt bth}: both models correct; {\tt nrm}: nonresponse model correct; {\tt ccm}: complete-case model correct; {\tt bad}: both models incorrect.\\
		**: Nominal level = 95\%.}
\end{table}

Results in Table 1 confirm our theoretical results, and clearly show that as
expected IPW has small bias in scenarios (1) and (2) only, PM has small bias
in scenarios (1) and (3), and DR has small bias in scenarios (1)-(3). In
scenario (4) where all models are incorrect, as expected all estimators are
significantly biased. When as in the first scenario, model misspecification is
absent, IPW has larger root mean squared error (RMSE)\ than PM, however DR is
comparable to PM, at least in this simulation setting. Interestingly, the RMSE
of DR follows closely  that of PM in scenarios (1) and (3) suggesting that the
potential efficiency loss incurred to obtain DR inference relative to PM
inference may not be substantial in practice. Table 1 of the Supplemental
Appendix summarizes simulation results assessing the performance of our
estimators of asymptotic variance and coverage of Wald confidence intervals
using estimated standard errors for the three estimators under consideration.
The results largely indicate that our standard error estimators are consistent
in all scenarios where the point estimators are also consistent, including
under partial model misspecification for the DR estimator (see comparison to
Monte Carlo standard errors in Table 1 of the Supplemental Appendix). However,
our standard error estimators appear to break down severely whenever model
mis-specification induces bias in parameter estimates. Interestingly, the
performance of the nonparametric bootstrap closely follows that of our
estimators in all instances and also appears to break down under bias inducing
model misspecification. We do not view this as a serious limitation given that
inferences are in such cases unreliable even with a consistent estimator of
standard error.

\subsection{A data application}

The empirical application concerns a study of the association between maternal
exposure to highly active antiretroviral therapy (HAART) during pregnancy and
birth outcomes among HIV-infected women in Botswana. A detailed description of
the study cohort has been presented elsewhere (Chen et al. 2012). The entire
study cohort consists of 33148 obstetrical records abstracted from 6 sites in
Botswana for 24 months. Our current analysis focuses on the subset of women
who were known to be HIV positive (n = 9711). The birth outcome of interest is
preterm delivery, defined as delivery
$<$
37 weeks gestation. 6.7\% of the outcomes are not observed. The data also
contain the following risk factors of interest that are also subject to
missingness (Table 2): whether CD4+ cell count is less than 200 cells/%
$\mu$%
L and whether a woman continued HAART from before pregnancy or not.%

	\begin{table}[ht]
		\centering
		\caption{Real data analysis: tabulation of missing data patterns. The total sample size is 9711. Missing variables are coded by 0. The first row represents the complete case}
				\vspace{10pt}
		\begin{tabular}{cccccccccccccc}
			\toprule
			Pattern (R) & Preterm Delivery & Low CD4 Count & Cont. HAART & percentage\\
			\midrule
			1 & 1 &  1 &  1  & 10.5\% \\
			2 &  0 &  1 &  1  & 0.7\% \\
			3  & 1  & 0  & 1  & 18.3\% \\
			4 &  0  & 0  & 1 &  1.6\% \\
			5 &  1 &  1 &  0  & 33.9\% \\
			6  & 0  & 1  & 0  & 1.5\% \\
			7 &  1  & 0  & 0  & 30.6\% \\
			8 &  0 &  0  & 0  & 2.9\% \\
			\bottomrule
		\end{tabular}
	\end{table}

Our goal is to correlate these factors with preterm delivery using a logistic
regression.\ In other words, the parameter of interest is the vector of
coefficients of the corresponding logistic regression.\ We implemented the
complete-case (CC)\ analysis together with three proposed estimators that
account for MNAR nonresponse: LDCM IPW, PM and DR estimators. Estimation of
the nonresponse process used the fairly generic specification$\ \log
\mathrm{Odds}_{r}\left(  L_{\left(  r\right)  };\alpha_{r}\right)  =\alpha
_{r}^{\prime}q_{r}\left(  L_{\left(  r\right)  }\right)  ,$ where
$q_{r}\left(  L_{\left(  r\right)  }\right)  $ included all main effects and
two-way interactions of components of $L_{\left(  r\right)  }$ while PM
specified the log-linear model $\Pr(L|R=1)\propto\exp\left\{  \mathbf{\eta
}^{\prime}L\right\}  .$%

	\begin{table}[ht]
		\centering
		\caption{Real data analysis: estimated odds ratios of preterm delivery associated with various risk factors.  The 95\% confidence intervals are estimated based on bootstrap samples}
		\vspace{10pt}
		\begin{tabular}{rccc}
			\toprule
			&  Low CD4 Count & Cont HAART \\
			\midrule
			CC & 0.782 (0.531,1.135) & 1.142 (0.810,1.620) \\
			IPW & 0.924 (0.631,1.338) & 1.180 (0.847,1.638) \\
			PM & 0.963 (0.704,1.318) & 1.175 (0.881,1.598) \\
			DR & 1.020 (0.742,1.397) & 1.158 (0.869,1.560) \\
			\bottomrule
		\end{tabular}
	\end{table}

Table 3 summarizes resuls for the complete analysis (CC) together with
Little's PM analysis and our two semiparametric estimators (IPW and\ DR). The
results suggest that the association between CD4 count and preterm delivery
may be subject to selection bias to a greater extent than that of HAART and
preterm delivery. In fact, the estimated odds ratio for CD4 count is about
20\% larger for IPW, PM\ and DR compare to the CC odds ratio, whereas the odds
ratio for HAART is quite similar for all four estimators. Although
PM\ generally appears less variable, there are no notable differences between
inferences obtained using IPW, PM or DR, providing no evidence that either IPW
or PM\ might be subject to misspecification bias.

\section{Inference for general DCM}

Consider a DCM with user-specified $F_{\varepsilon},$ a well-defined
continuous CDF. Local identification under assumption $\left(  \ref{ID}%
\right)  $ is best understood with discrete data. In this vein, suppose that
$L_{\left(  r\right)  }$ takes on $M_{r}$ levels, then $\Delta\mu_{1r}\left(
L_{(r)}\right)  $ depends at most on $M_{r}$ unknown parameters. However, note
that for user-supplied $M_{r}$-dimensional function $G_{r}=g_{r}\left(
L_{\left(  r\right)  }\right)  .$ Let $W_{r}\left(  G_{r}\right)  =G_{r}%
\times\left[  1\left\{  R=r\right\}  -1\left\{  R=1\right\}  \Pi_{r}/\Pi
_{1}\right]  .$ It is straightforward to verify that
\begin{equation}
E\left\{  W_{r}\left(  G_{r}\right)  \right\}  =0\text{ for }%
r=2,...\label{Moment}%
\end{equation}
yielding the $M_{r}$ restrictions needed to identify each $\Delta\mu_{1r}.$
Naturally, components of $G_{r}$ should be chosen appropriately to avoid
redundancy and linear dependence. A similar argument could in principle be
carefully crafted to establish local identification if $L$ contains continuous
components. However, this is not further pursued in this paper. Interestingly,
equation $\left(  \ref{Moment}\right)  $ motivates a simple approach for
estimating $\Pi_{r}$ in practice. Suppose that one posits a parametric model
$\Delta\mu_{1r}\left(  L_{\left(  r\right)  };\alpha_{r}\right)  $ for
$\Delta\mu_{1r}\left(  L_{\left(  r\right)  }\right)  $ with finite
dimensional unknown parameter $\alpha_{r},$ for all $r.$ Then, the following
empirical version of $\left(  \ref{Moment}\right)  $ would in principle
deliver an estimator $\widehat{\mathbf{\alpha}}\mathbf{=}\left\{
\widehat{\alpha}_{r}:r\right\}  $ of $\mathbf{\alpha=}\left\{  \alpha
_{r}:r\right\}  .$%
\[
\mathbb{P}_{n}\left\{  W_{r}\left(  \widehat{G}_{r};\widehat{\mathbf{\alpha}%
}\right)  \right\}  =0\text{ for }r=2,...
\]
where $W_{r}\left(  \widehat{G}_{r};\widehat{\mathbf{\alpha}}\right)
=\widehat{G}_{r}\times\left[  1\left\{  R=r\right\}  -1\left\{  R=1\right\}
\Pi_{r}\left(  \widehat{\mathbf{\alpha}}\right)  /\Pi_{1}\left(
\widehat{\mathbf{\alpha}}\right)  \right]  .$ A convenient choice for
$\widehat{G}_{r}=\partial\Delta\mu_{1r}\left(  L_{\left(  r\right)
};\widehat{\alpha}_{r}\right)  /\partial\widehat{\alpha}_{r}.$ Under mild
regularity conditions, $\widehat{\mathbf{\alpha}}$ will be consistent and
asymptotically normal provided that $\Delta\mu_{1r}\left(  L_{\left(
	r\right)  };\alpha_{r}\right)  $ is correctly specified for all $r$.

Given a consistent estimator of $\Pi_{1},$ IPW inferences about $\beta_{0}$
may be obtained as described in previous sections. Likewise, maximum
likelihood estimation is straightforward by maximizing a model for the
likelihood given in Lemma 1$.$ Unfortunately, outside of the LDCM, to the best
of our knowledge, it does not appear possible to obtain DR and MR inferences
for DCMs$.$

The above analysis requires evaluation of the integral defining $\Pi_{r}%
.$Thus, let
\[
Q_{r}\left(  \varepsilon\right)  =%
{\displaystyle\prod\limits_{s\neq r}}
F_{\varepsilon}\left(  \Delta\mu_{1s}\left(  L_{\left(  s\right)  }\right)
-\Delta\mu_{1r}\left(  L_{\left(  r\right)  }\right)  +\varepsilon\right)  .
\]
A reliable approximation of $\Pi_{r}=\int Q_{r}\left(  \varepsilon\right)
f_{\varepsilon}\left(  \varepsilon\right)  d\varepsilon$ can effectively be
achieved numerically by Gauss-Hermite Quadrature (Liu and Pierce, 1994). For
instance, suppose that $f_{\varepsilon}$ is standard normal, then the
approximate Gaussian Discrete Choice Model is given by $\Pi_{r}\approx
\sum_{m=1}^{M}Q_{r}\left(  \varepsilon_{m}\right)  w_{m},$ where the nodes
$\varepsilon_{m}$ are the zeroes of the mth order Hermite polynomial and
$w_{m}$ are suitably defined weights (Davis \& Rabinowitz, 1975)

\section{{\protect\Large Conclusion}}

\noindent In this paper, we have described the DCM as an all-purpose, flexible
and easy-to-implement general class of models for nonmonotone nonignorable
nonresponse. \ The LDCM\ has several advantages including giving rise to four
distinct strategies for inference: IPW, PM, DR and MR estimation. \ Simulation
studies and an application suggest good finite sample performance of IPW, PM
and DR estimation; although not directly evaluated, we expect the same to
apply to MR estimation.

Identification conditions such as CCMV are not empirically testable and
therefore, it is important that inferences are assessed for sensitivity to
violation of such assumptions. Such an approach for sensitivity analysis for
violation of CCMV restriction is outlined in the Supplemental Appendix.

\clearpage

\begin{center}
	
	{\LARGE Supplemental Appendix}
	
	{\LARGE Semiparametric inference under a discrete choice}
	
	{\LARGE model for nonmonotone missing not at random data }
	
	{\Large Eric J. Tchetgen Tchetgen, Linbo Wang, BaoLuo Sun } $\ $
	
	{\large Department of Biostatistics}$,$
	
	{\large Harvard University}
\end{center}
\setcounter{equation}{0}
\setcounter{figure}{0}
\setcounter{table}{0}
\setcounter{page}{1}
\makeatletter
\renewcommand{\theequation}{S\arabic{equation}}
\renewcommand{\thefigure}{S\arabic{figure}}
\renewcommand{\thetable}{S\arabic{table}}
\setcounter{section}{0}

\section{Sensitivity analysis for CCMV}

Identification conditions such as CCMV are not generally empirically testable
and therefore, it is important that inferences in a given analysis are
assessed for sensitivity to violation of such assumptions. Specifically, a
violation of the CCMV assumption can occur if for some $r,$%
\[
R\not \perp \!\!\!\perp L_{(-r)}|L_{(r)},R\in\left\{  1,r\right\}  ,
\]
which can be encoded by specifying the degree of departure from the
identifying assumption, on the odds ratio scale using the selection bias
function:%
\[
\theta_{r}\left(  L_{\left(  -r\right)  },L_{(r)}\right)  =\frac{\pi
	_{r}\left(  L_{(r)},L_{\left(  -r\right)  }\right)  \pi_{1}\left(
	L_{(r)},L_{\left(  -r\right)  }=0\right)  }{\pi_{1}\left(  L_{(r)},L_{\left(
		-r\right)  }\right)  \pi_{r}\left(  L_{(r)},L_{\left(  -r\right)  }=0\right)
}.
\]
CCMV corresponds to the null $\theta_{r}\left(  L_{\left(  -r\right)
},L_{(r)}\right)  =1$ for all $r,$ and $\theta_{r}\left(  L_{\left(
-r\right)  },L_{(r)}\right)  \not =1$ for some $r$ indicates violation of the
assumption. The function $\theta_{r}\left(  \cdot,\cdot\right)  $ is not
nonparametrically identified from the observed data. Therefore we propose that
one may specify a functional form for $\theta_{r}\left(  \cdot,\cdot\right)  $
for use in a sensitivity analysis in the spirit of Robins et al (1999).
Hereafter, suppose that one has specified functions $\mathbf{\theta=}\left\{
\theta_{r}:r\right\}  .$ For such specification, we describe IPW, PM and DR
estimation incorporating a non-null $\theta_{r}$.

\ For IPW estimation, we propose to modify $W_{r}$\ of Section 5 as follows.
Let $W_{r}\left(  G_{r};\mathbf{\alpha,}\theta_{r}\right)  =G_{r}\times\left[
1\left\{  R=r\right\}  -1\left\{  R=1\right\}  \theta_{r}\left(  L\right)
\Pi_{r}\left(  \mathbf{\alpha}\right)  /\Pi_{1}\left(  \mathbf{\alpha}\right)
\right]  ,$ and denote by $\widehat{\mathbf{\alpha}}\left(  \mathbf{\theta
}\right)  $ the solution to $\mathbb{P}_{n}W_{r}\left(  G_{r}%
;\widehat{\mathbf{\alpha}}\left(  \mathbf{\theta}\right)  \mathbf{,}\theta
_{r}\right)  =0,$ then a consistent IPW\ estimator $\widehat{\beta}%
_{ipw}\left(  \mathbf{\theta}\right)  $ solves equation $\left(  10\right)  $
in the main text with $\Pi_{1}$ replaced by $\Pi_{1}^{\ast}\left(
\widehat{\mathbf{\alpha}}\left(  \mathbf{\theta}\right)  \right)  =\left\{
1+\sum_{r\neq1}\theta_{r}\left(  L\right)  \Pi_{r}\left(
\widehat{\mathbf{\alpha}}\left(  \mathbf{\theta}\right)  \right)  /\Pi
_{1}\left(  \widehat{\mathbf{\alpha}}\left(  \mathbf{\theta}\right)  \right)
\right\}  ^{-1}$

Likewise, PM estimation hinges on the following expression
\begin{align*}
&  \bigskip E\left\{  U(L;\beta)|R=r,L_{\left(  r\right)  };\widetilde{\eta
},\mathbf{\theta}\right\} \\
&  =\frac{\int\theta_{r}\left(  L_{\left(  -r\right)  },L_{(r)}\right)
	U(l_{\left(  -r\right)  },L_{\left(  r)\right)  };\beta)f\left(  l_{\left(
		-r\right)  },L_{\left(  r)\right)  }|R=1;\mathbf{\eta}\right)  d\mu\left(
	l_{\left(  -r\right)  }\right)  }{\int\theta_{r}\left(  L_{\left(  -r\right)
	},L_{(r)}\right)  f\left(  l_{\left(  -r\right)  },L_{\left(  r)\right)
}|R=1;\mathbf{\eta}\right)  d\mu\left(  l_{\left(  -r\right)  }\right)  }%
\end{align*}
which may be used in place of $E\left\{  U(L;\beta)|R=1,L_{\left(  r\right)
};\widetilde{\eta}\right\}  $ in equation $\left(  12\right)  ,$ which in turn
may be used to obtain\ the PM estimator $\hat{\beta}_{pm}\left(
\mathbf{\theta}\right)  $. Finally, for a given value of $\mathbf{\theta,}$
the DR estimator $\hat{\beta}_{dr}\left(  \mathbf{\theta}\right)  $ solves
equation $\left(  14\right)  $ with $V\left(  \widehat{\beta}_{dr}%
,\widetilde{\mathbf{\alpha}}\mathbf{,}\widetilde{\mathbf{\eta}}\right)  $
replaced by
\begin{align*}
V\left(  \beta,\widehat{\mathbf{\alpha}}\left(  \mathbf{\theta}\right)
\mathbf{,\widetilde{\eta};\theta}\right)   &  =\left\{  \frac{1\left(
	R=1\right)  }{\Pi_{1}^{\ast}\left(  \widehat{\mathbf{\alpha}}\left(
	\mathbf{\theta}\right)  \right)  }U(L;\beta)\right\} \\
&  -\frac{1\left(  R=1\right)  }{\Pi_{1}^{\ast}\left(  \widehat{\mathbf{\alpha
		}}\left(  \mathbf{\theta}\right)  \right)  }\sum_{r\pm1}\Pi_{r}^{\ast}\left(
	\widehat{\mathbf{\alpha}}\left(  \mathbf{\theta}\right)  \right)  E\left[
	U(L;\beta)|L_{\left(  r\right)  },R=r;\widetilde{\eta},\mathbf{\theta}\right]
	\\
	&  +\sum_{r\pm1}I\left(  R=r\right)  E\left[  U(L;\beta)|L_{\left(  r\right)
	},R=r;\widetilde{\eta},\mathbf{\theta}\right]  ,
	\end{align*}
	where
	\[
	\Pi_{r}^{\ast}\left(  \widehat{\mathbf{\alpha}}\left(  \mathbf{\theta}\right)
	\right)  =\frac{\theta_{r}\left(  L\right)  \Pi_{r}\left(
		\widehat{\mathbf{\alpha}}\left(  \mathbf{\theta}\right)  \right)  /\Pi
		_{1}\left(  \widehat{\mathbf{\alpha}}\left(  \mathbf{\theta}\right)  \right)
	}{\left\{  1+\sum_{r^{\prime}\neq1}\theta_{r^{\prime}}\left(  L\right)
	\Pi_{r^{\prime}}\left(  \widehat{\mathbf{\alpha}}\left(  \mathbf{\theta
	}\right)  \right)  /\Pi_{1}\left(  \widehat{\mathbf{\alpha}}\left(
	\mathbf{\theta}\right)  \right)  \right\}  },
\]

A sensitivity analysis then entails reporting $\hat{\beta}_{ipw}\left(
\mathbf{\theta}\right)  $, $\hat{\beta}_{pm}\left(  \mathbf{\theta}\right)  $
or $\hat{\beta}_{dr}\left(  \mathbf{\theta}\right)  $ for a range of values of
$\mathbf{\theta.}$

\section{Proof of Lemmas}

\noindent\textbf{Proof of Lemma 1: }The result follows from the following
generalized odds ratio representation of the joint likelihood of $f(R,L)$ (see
Chen, 2007 and Tchetgen Tchetgen et al, 2010)
\[
f(R,L)=\frac{f\left(  R|L=0\right)  f(L|R=1)\mathrm{OR}\left(  R,L\right)  }{%
	{\displaystyle\iint}
	f\left(  r^{\ast}|L=0\right)  f(l^{\ast}|R=1)\mathrm{OR}\left(  r^{\ast
	},l^{\ast}\right)  d\mu\left(  r^{\ast},l^{\ast}\right)  },
\]
provided that $%
{\displaystyle\iint}
f\left(  r^{\ast}|L=0\right)  f(l^{\ast}|R=1)\mathrm{OR}\left(  r^{\ast
},l^{\ast}\right)  d\mu\left(  r^{\ast},l^{\ast}\right)  <\infty,$ where the
generalized odds ratio function $\mathrm{OR}\left(  R,L\right)  $ is defined
as
\[
\mathrm{OR}\left(  R,L\right)  =\frac{f\left(  R,L\right)  f(R=1,L=0)}%
{f\left(  R=1,L\right)  f(R,L=0)}.
\]
Then
\begin{align*}
&  \frac{f\left(  R|L=0\right)  f(L|R=1)\mathrm{OR}\left(  R,L\right)  }{%
	{\displaystyle\iint}
	f\left(  r^{\ast}|L=0\right)  f(l^{\ast}|R=1)\mathrm{OR}\left(  r^{\ast
	},l^{\ast}\right)  d\mu\left(  r^{\ast},l^{\ast}\right)  }\\
&  =\frac{\frac{f\left(  R|L=0\right)  }{f\left(  R=1|L=0\right)  }%
	\mathrm{OR}\left(  R,L\right)  f(L|R=1)}{%
	{\displaystyle\iint}
	\frac{f\left(  r^{\ast}|L=0\right)  }{f\left(  R=1|L=0\right)  }%
	\mathrm{OR}\left(  r^{\ast},l^{\ast}\right)  f(l^{\ast}|R=1)d\mu\left(
	r^{\ast},l^{\ast}\right)  }\\
&  =\frac{%
	{\displaystyle\prod\limits_{r\neq1}}
	\mathrm{Odds}_{r}\left(  L\right)  ^{I\left(  R=r\right)  }f\left(
	L|R=1\right)  f(L|R=1)}{%
	{\displaystyle\iint}
	{\displaystyle\prod\limits_{r\neq1}}
	\mathrm{Odds}_{r}\left(  l^{\ast}\right)  ^{I\left(  r^{\ast}=r\right)
	}f\left(  l^{\ast}|R=1\right)  d\mu\left(  r^{\ast},l^{\ast}\right)  }%
\end{align*}
proving the result.

\noindent\textbf{Proof of Lemma 2: }The complete-case joint distribution
$f(L|R=1)$ is nonparametrically just-identified under assumption (1).
Furthermore, pairwise MAR implies that $\mathrm{Odds}_{r}\left(  L\right)
=\mathrm{Odds}_{r}\left(  L_{\left(  r\right)  }\right)  $ is
nonparametrically just-identified from data $\left\{  (R,L_{(R)}):R\in\left\{
1,r\right\}  \right\}  ,$ because $L_{(-r)}$ is MAR conditional on $L_{\left(
	R\right)  }$ and $R\in\left\{  1,r\right\}  .$ Specifically,
\begin{align*}
&  \Pr\left\{  R=r|L,R\in\left\{  1,r\right\}  \right\} \\
&  =\frac{\Pr\left\{  R=r,L\right\}  }{\Pr\left\{  L,R\in\left\{  1,r\right\}
	\right\}  }\\
&  =\frac{\mathrm{Odds}_{r}\left(  L_{\left(  r\right)  }\right)  f\left(
	L|R=1\right)  f(L|R=1)}{\mathrm{Odds}_{r}\left(  L_{\left(  r\right)
	}\right)  f\left(  L|R=1\right)  f(L|R=1)+f\left(  L|R=1\right)  f(L|R=1)}\\
&  =\frac{\mathrm{Odds}_{r}\left(  L_{\left(  r\right)  }\right)
}{\mathrm{Odds}_{r}\left(  L_{\left(  r\right)  }\right)  +1},\text{ }%
\end{align*}
proving the result.

\noindent\textbf{Proof of Theorem 3: }The result essentially follows from the
following DR\ property of $V\left(  \beta,\mathbf{\alpha,\eta}\right)  .$ Let
$V\left(  \beta,\mathbf{\alpha}^{\ast}\mathbf{,\eta}_{0}\right)  $ denote the
estimating function evaluated at the incorrect $\Pi_{r}$ and true
$E[U(L;\beta)|L_{\left(  r\right)  },R=1]$ for all $r.$ Likewise let $V\left(
\beta,\mathbf{\alpha}_{0}\mathbf{,\eta}^{\ast}\right)  $ for the opposite
setting. DR property holds if $E\left\{  V\left(  \beta_{0},\mathbf{\alpha
}^{\ast}\mathbf{,\eta}_{0}\right)  \right\}  =\ E\left\{  V\left(  \beta
_{0},\mathbf{\alpha}_{0}\mathbf{,\eta}^{\ast}\right)  \right\}  =0.$ First,
note that under $\mathcal{M}_{R},$ $\widetilde{\mathbf{\alpha}}\rightarrow
\mathbf{\alpha}_{0}$ and $\widetilde{\mathbf{\eta}}\rightarrow\mathbf{\eta
}^{\ast}$ in probability$,$ then $\mathbb{P}_{n}V\left(  \beta_{0}%
,\widetilde{\mathbf{\alpha}}\mathbf{,}\widetilde{\mathbf{\eta}}\right)
\rightarrow E\left\{  V\left(  \beta_{0},\mathbf{\alpha}_{0}\mathbf{,\eta
}^{\ast}\right)  \right\}  $ in probability by Continuous Mapping Theorem and
the Law of Large Numbers. We also have that%

\begin{align*}
E\left(  V\left(  \beta,\mathbf{\alpha}_{0}\mathbf{,\eta}^{\ast}\right)
\right)   &  =E\left\{  \frac{1\left(  R=1\right)  }{\Pi_{1}\left(
	\mathbf{\alpha}_{0}\right)  }U(L;\beta_{0})\right. \\
&  \left.  -\sum_{r\neq1}\left(  \frac{1\left(  R=1\right)  \Pi_{r}\left(
	\mathbf{\alpha}_{0}\right)  }{\Pi_{1}\left(  \mathbf{\alpha}\right)
}-1\left(  R=r\right)  \right)  E\left[  U(L;\beta)|L_{\left(  r\right)
},R=1;\mathbf{\eta}^{\ast}\right]  \right\} \\
&  =E\left\{  \frac{E\left\{  1\left(  R=1\right)  |L\right\}  }{\Pi
	_{1}\left(  \mathbf{\alpha}_{0}\right)  }U(L;\beta_{0})\right. \\
&  -\sum_{r\neq1}\underset{=0}{\underbrace{\left(  \frac{E\left\{  1\left(
			R=1\right)  |L\right\}  \Pi_{r}\left(  \mathbf{\alpha}_{0}\right)  }{\Pi
			_{1}\left(  \mathbf{\alpha}\right)  }-E\left\{  1\left(  R=r\right)
		|L\right\}  \right)  }}E\left[  U(L;\beta)|L_{\left(  r\right)  }%
,R=1;\mathbf{\eta}^{\ast}\right] \\
&  =E\left[  U(L;\beta_{0})\right]  =0
\end{align*}
By the same token, under $\mathcal{M}_{L}$, $\widetilde{\mathbf{\alpha}%
}\rightarrow\mathbf{\alpha}^{\ast}$ and $\widetilde{\mathbf{\eta}}%
\rightarrow\mathbf{\eta}_{0}$ in probability$,$ then $\mathbb{P}_{n}V\left(
\beta_{0},\widetilde{\mathbf{\alpha}}\mathbf{,}\widetilde{\mathbf{\eta}%
}\right)  \rightarrow E\left\{  V\left(  \beta_{0},\mathbf{\alpha}^{\ast
}\mathbf{,\eta}_{0}\right)  \right\}  .$ Next we show that $E\left\{  V\left(
\beta_{0},\mathbf{\alpha}^{\ast}\mathbf{,\eta}_{0}\right)  \right\}  =0.$ Note
that for all $\alpha$
\begin{align*}
\frac{1}{\Pi_{1}\left(  \mathbf{\alpha}\right)  }  &  =1+\sum_{r\neq1}%
\frac{\Pi_{r}\left(  \mathbf{\alpha}\right)  }{\Pi_{1}\left(  \mathbf{\alpha
	}\right)  }\\
&  =1+\sum_{r\neq1}\mathrm{Odds}_{r}\left(  L_{\left(  r\right)
};\mathbf{\alpha}\right)  .
\end{align*}
Then we have that
\begin{align*}
E\left(  V\left(  \beta,\mathbf{\alpha}_{0}\mathbf{,\eta}^{\ast}\right)
\right)   &  =E\left\{  \frac{1\left(  R=1\right)  }{\Pi_{1}\left(
	\mathbf{\alpha}^{\ast}\right)  }\left\{  U(L;\beta_{0})-\sum_{r\neq1}\Pi
_{r}\left(  \mathbf{\alpha}^{\ast}\right)  E\left[  U(L;\beta)|L_{\left(
	r\right)  },R=1;\mathbf{\eta}_{0}\right]  \right\}  \right. \\
&  \left.  +\sum_{r\neq1}1\left(  R=r\right)  E\left[  U(L;\beta)|L_{\left(
	r\right)  },R=1;\mathbf{\eta}_{0}\right]  \right\} \\
&  =E\left\{  1\left(  R=1\right)  \left\{  \frac{U(L;\beta_{0})}{\Pi
	_{1}\left(  \mathbf{\alpha}^{\ast}\right)  }-\sum_{r\neq1}\frac{\Pi_{r}\left(
	\mathbf{\alpha}^{\ast}\right)  }{\Pi_{1}\left(  \mathbf{\alpha}^{\ast}\right)
}E\left[  U(L;\beta)|L_{\left(  r\right)  },R=1;\mathbf{\eta}_{0}\right]
\right\}  \right. \\
&  \left.  +\sum_{r\neq1}1\left(  R=r\right)  E\left[  U(L;\beta)|L_{\left(
	r\right)  },R=1;\mathbf{\eta}_{0}\right]  \right\} \\
&  =E\left\{  \underset{=0}{\underbrace{\sum_{r\neq1}\mathrm{Odds}_{r}\left(
		L_{\left(  r\right)  };\mathbf{\alpha}^{\ast}\right)  \left(  E\left[
		U(L;\beta_{0})|R=1,L_{\left(  r\right)  }\right]  -E\left[  U(L;\beta
		)|L_{\left(  r\right)  },R=1;\mathbf{\eta}_{0}\right]  \right)  }}\right\} \\
&  \left.  1\left(  R=1\right)  U(L;\beta_{0})+\sum_{r\neq1}1\left(
R=r\right)  E\left[  U(L;\beta)|L_{\left(  r\right)  },R=1;\mathbf{\eta}%
_{0}\right]  \right\} \\
&  =E\left\{  1\left(  R=1\right)  U(L;\beta_{0})+\sum_{r\neq1}1\left(
R=r\right)  E\left[  U(L;\beta)|L_{\left(  r\right)  },R=1;\mathbf{\eta}%
_{0}\right]  \right\} \\
&  =E\left\{  1\left(  R=1\right)  U(L;\beta_{0})+\sum_{r\neq1}1\left(
R=r\right)  E\left[  U(L;\beta)|L_{\left(  r\right)  },R=r\right]  \right\} \\
&  =E\left\{  1\left(  R=1\right)  E[U(L;\beta_{0})|R=1]+\sum_{r\neq1}1\left(
R=r\right)  E\left[  U(L;\beta)|R=r\right]  \right\} \\
&  =E[U(L;\beta_{0})]=0
\end{align*}
proving the result.\noindent

\noindent\textbf{Proof of Corollary 4: }$E\left(  V\left(  \beta
,\mathbf{\alpha,\eta}\right)  \right)  $\textbf{ }can be written%
\begin{align*}
&  E\left(  V\left(  \beta,\mathbf{\alpha,\eta}\right)  \right) \\
&  =E\left\{  \sum_{r\neq1}\frac{1\left(  R=1\right)  \Pi_{r}\left(
	\mathbf{\alpha}\right)  }{\Pi_{1}\left(  \mathbf{\alpha}\right)  }%
U(L;\beta_{0})-\sum_{r\neq1}\frac{1\left(  R=1\right)  \Pi_{r}\left(
	\mathbf{\alpha}\right)  }{\Pi_{1}\left(  \mathbf{\alpha}\right)  }E\left[
U(L;\beta)|L_{\left(  r\right)  },R=1;\mathbf{\eta}\right]  \right. \\
&  \left.  +\sum_{r\neq1}1\left(  R=r\right)  E\left[  U(L;\beta)|L_{\left(
	r\right)  },R=1;\mathbf{\eta}\right]  +1\left(  R=1\right)  U(L;\beta
_{0})\right\} \\
&  =E\left\{  \sum_{r\neq1}\frac{1\left(  R=1\right)  \Pi_{r}\left(
	\mathbf{\alpha}\right)  }{\Pi_{1}\left(  \mathbf{\alpha}\right)  }%
U(L;\beta_{0})-\sum_{r\neq1}\frac{1\left(  R=1\right)  \Pi_{r}\left(
	\mathbf{\alpha}\right)  }{\Pi_{1}\left(  \mathbf{\alpha}\right)  }E\left[
U(L;\beta)|L_{\left(  r\right)  },R=1;\mathbf{\eta}\right]  \right. \\
&  \left.  +\sum_{r\neq1}1\left(  R=r\right)  \left\{  E\left[  U(L;\beta
)|L_{\left(  r\right)  },R=1;\mathbf{\eta}\right]  -U(L;\beta_{0})\right\}
+U(L;\beta_{0})\right\} \\
&  =E\left[  \sum_{r\neq1}\left\{  1\left(  R=1\right)  \mathrm{Odds}%
_{r}\left(  L_{\left(  r\right)  };\mathbf{\alpha}\right)  -1\left(
R=r\right)  \right\}  \left\{  U(L;\beta_{0})-E\left[  U(L;\beta)|L_{\left(
	r\right)  },R=1;\mathbf{\eta}\right]  \right\}  \right]
\end{align*}
Under $\mathcal{M}_{R}\left(  r\right)  ,$ we have that $\mathrm{Odds}%
_{r}\left(  L_{\left(  r\right)  };\widetilde{\mathbf{\alpha}}\right)
\rightarrow\mathrm{Odds}_{r}\left(  L_{\left(  r\right)  };\mathbf{\alpha}%
_{0}\right)  $ in probability, and
\begin{align*}
&  E\left[  \left\{  1\left(  R=1\right)  \mathrm{Odds}_{r}\left(  L_{\left(
	r\right)  };\mathbf{\alpha}_{0}\right)  -1\left(  R=r\right)  \right\}
\left\{  U(L;\beta_{0})-E\left[  U(L;\beta)|L_{\left(  r\right)
},R=1;\mathbf{\eta}^{\ast}\right]  \right\}  \right] \\
&  =E\left[  \left\{  1\left(  R=1\right)  \frac{\Pi_{r}}{\Pi_{1}}-1\left(
R=r\right)  \right\}  \left\{  U(L;\beta_{0})-E\left[  U(L;\beta)|L_{\left(
	r\right)  },R=1;\mathbf{\eta}^{\ast}\right]  \right\}  \right] \\
&  =E\left[  \left\{  \Pi_{r}-E\left[  1\left(  R=r\right)  |L\right]
\right\}  \left\{  U(L;\beta_{0})-E\left[  U(L;\beta)|L_{\left(  r\right)
},R=1;\mathbf{\eta}^{\ast}\right]  \right\}  \right] \\
&  =0
\end{align*}
Likewise, under $\mathcal{M}_{L}\left(  r\right)  ,$ we have that $E\left[
U(L;\beta)|L_{\left(  r\right)  },R=1;\widetilde{\mathbf{\eta}}\right]
\rightarrow E\left[  U(L;\beta)|L_{\left(  r\right)  },R=1;\mathbf{\eta}%
_{0}\right]  $ in probability, and%

\begin{align*}
&  E\left[  \left\{  1\left(  R=1\right)  \mathrm{Odds}_{r}\left(  L_{\left(
	r\right)  };\mathbf{\alpha}^{\ast}\right)  -1\left(  R=r\right)  \right\}
\left\{  U(L;\beta_{0})-E\left[  U(L;\beta)|L_{\left(  r\right)
},R=1;\mathbf{\eta}_{0}\right]  \right\}  \right] \\
&  =E\left[  1\left(  R=1\right)  \mathrm{Odds}_{r}\left(  L_{\left(
	r\right)  };\mathbf{\alpha}^{\ast}\right)  \left\{  E\left\{  U(L;\beta
_{0})|R=1,L_{(r)}\right\}  -E\left[  U(L;\beta)|L_{\left(  r\right)
},R=1;\mathbf{\eta}_{0}\right]  \right\}  \right] \\
&  -E\left[  \left\{  1\left(  R=r\right)  \right\}  \left\{  E\left\{
U(L;\beta_{0})|R=r,L_{(r)}\right\}  -E\left[  U(L;\beta)|L_{\left(  r\right)
},R=1;\mathbf{\eta}_{0}\right]  \right\}  \right] \\
&  =-E\left[  \left\{  1\left(  R=r\right)  \right\}  \left\{  E\left\{
U(L;\beta_{0})|R=1,L_{(r)}\right\}  -E\left[  U(L;\beta)|L_{\left(  r\right)
},R=1;\mathbf{\eta}_{0}\right]  \right\}  \right] \\
&  =0
\end{align*}
proving the result.

\pagebreak

\section{\bigskip Additional Simulation Results}

Table \ref{tab:est1} shows Monte Carlo results comparing the proposed large sample
estimator of standard deviation (and corresponding coverage probabilities of
Wald 95\% confidence intervals) of IPW, PM and DR estimators of $\beta$ to
corresponding Monte Carlo standard deviations .%

\begin{table}
	\begin{center}
		\caption{Monte Carlo results of the IPW, PM and DR estimators: bias, standard error and root mean squared error. The true value of $\beta$  is 0.634, and the sample size is 2000.} 
	\vspace*{10pt}
		\label{tab:est1}
		\begin{tabular}{rccccccccccccccc}
			\toprule
			& \multicolumn{1}{c}{\tt bth$^*$} & \multicolumn{1}{c}{\tt nrm} &   \multicolumn{1}{c}{\tt ccm} & \multicolumn{1}{c}{\tt bad} \\
			\midrule
			\multicolumn{1}{l}{Bias(SE) \quad \quad}	 & & & \\[3pt]
			IPW & -0.004(0.002) & -0.004(0.002) & -0.641(0.012) & -0.641(0.012) \\ 
			PM & -0.002(0.001) & -0.367(0.002) & -0.002(0.001) & -0.367(0.002) \\ 
			DR & -0.002(0.002) & -0.006(0.002) & -0.002(0.002) & -0.371(0.003) \\ [12pt]
			\multicolumn{1}{l}{RMSE \quad \quad}	 & & & \\[3pt]
			IPW & 0.072 & 0.072 & 0.748 & 0.748 \\ 
			PM & 0.046 & 0.373 & 0.046 & 0.373 \\ 
			DR & 0.048 & 0.057 & 0.057 & 0.385 \\ 
			\bottomrule
		\end{tabular}
	\end{center}	
	\footnotesize{*: {\tt bth}: both models correct; {\tt nrm}: nonresponse model correct; {\tt ccm}: complete-case model correct; {\tt bad}: both models incorrect.}
\end{table}

\bigskip

\pagebreak

\end{document}